\documentclass[sigconf,natbib,screen]{acmart}

\acmSubmissionID{1423}

\usepackage{acronym}
\usepackage[skip=0pt]{caption}
\usepackage[inline]{enumitem}
\usepackage{booktabs}
\usepackage{makecell}
\usepackage{multirow}
\usepackage{subcaption}

\AtBeginDocument{%
  \providecommand\BibTeX{{%
    \normalfont B\kern-0.5em{\scshape i\kern-0.25em b}\kern-0.8em\TeX}}}

\acrodef{DS}{dialogue system}
\acrodef{TDS}{task-oriented dialogue system}
\acrodef{ReDial}{recommendation dialogues}
\acrodef{IR}{information retrieval}
\acrodef{AMT}{Amazon Mechanical Turk}
\acrodef{SAT}{satisfied}
\acrodef{DSAT}{dissatisfied}

\newcommand{\header}[1]{\vspace{1mm}\noindent\textbf{#1.}}

\copyrightyear{2022}
\acmYear{2022}
\setcopyright{none}\acmConference[SIGIR '22]{Proceedings of the 45th International ACM SIGIR Conference on Research and Development in Information Retrieval}{July 11--15, 2022}{Madrid, Spain}
\acmBooktitle{Proceedings of the 45th International ACM SIGIR Conference on Research and Development in Information Retrieval (SIGIR '22), July 11--15, 2022, Madrid, Spain}
\acmPrice{15.00}
\acmDOI{10.1145/3477495.3531798}
\acmISBN{978-1-4503-8732-3/22/07}

\author{Clemencia Siro}
\affiliation{%
  \institution{University of Amsterdam}
  \city{Amsterdam}
  \country{The Netherlands}}
\email{c.n.siro@uva.nl}

\author{Mohammad Aliannejadi}
\affiliation{%
  \institution{University of Amsterdam}
  \city{Amsterdam}
\country{The Netherlands}}
\email{m.aliannejadi@uva.nl}

\author{Maarten de Rijke}
\affiliation{%
  \institution{
  University of Amsterdam}
  \city{Amsterdam}
  \country{The Netherlands}}%
  
\email{m.derijke@uva.nl}

\allowdisplaybreaks

\begin{document}

\title[Understanding User Satisfaction with Task-oriented Dialogue Systems]{Understanding User Satisfaction \\ with Task-oriented Dialogue Systems}

\begin{abstract}
\Acp{DS} are evaluated depending on their type and purpose. 
Two categories are often distinguished: 
\begin{enumerate*} 
\item \acp{TDS}, which are typically evaluated on utility, i.e., their ability to complete a specified task, and 
\item open-domain chat-bots, which are evaluated on the user experience, i.e., based on their ability to engage a person. 
\end{enumerate*} 
What is the influence of \emph{user experience} on the user satisfaction rating of \acp{TDS} as opposed to, or in addition to, \emph{utility}? 
We collect data by providing an additional annotation layer for dialogues sampled from the ReDial dataset, a widely used conversational recommendation dataset. 
Unlike prior work, we annotate the sampled dialogues at both the turn and dialogue level on six dialogue aspects: \emph{relevance}, \emph{interestingness}, \emph{understanding}, \emph{task completion}, \emph{efficiency}, and \emph{interest arousal}. 
The annotations allow us to study how different dialogue aspects influence user satisfaction. We introduce a comprehensive set of user experience aspects derived from the annotators’ open comments that can influence users' overall impression.
We find that the concept of satisfaction varies across annotators and dialogues, and show that  a relevant turn is significant for some annotators, while for others, an interesting turn is all they need.
Our analysis indicates that the proposed user experience aspects provide a fine-grained analysis of user satisfaction that is not captured by a monolithic overall human rating.

\end{abstract}



\maketitle

\acresetall

\section{Introduction}
 
Recent research into the evaluation of conversational systems such as dialogue systems and conversational recommender systems has proposed automatic metrics that are meant to correlate well with human judgements~\citep{Deriu2020SurveyOE}.
Many of these standard evaluation metrics have been shown to be ineffective in dialogue evaluation \citep{liu-etal-2016-evaluate, Deriu2020SurveyOE}. 
As a consequence, a significant amount of dialogue research relies on human evaluation to measure a system's effectiveness. 
Recently, estimating a user's overall satisfaction with system interaction has gained momentum as the core evaluation metric for \ac{TDS}~\citep{10.1145/2854946.2854961,10.1145/3269206.3271802}. Though useful and effective, overall user satisfaction does not necessarily give insights on what aspect or dimensions the \ac{TDS} is performing well.
Knowing why a user is satisfied or dissatisfied helps the conversational system recover from an error and optimise toward an individual aspect to avoid total dissatisfaction during an interaction session.

Understanding \emph{user satisfaction} with a \ac{TDS} at a fine-grained level is vital at both the design and evaluation stages. 
Metrics such as engagement, relevance, and interestingness have been investigated to understand fine-grained user satisfaction and how they determine a user's overall satisfaction in different scenarios and applications \citep{Venkatesh2018OnEA,gopalakrishnan19_interspeech,see-etal-2019-makes}. For \ac{TDS}, user satisfaction is modelled as an evaluation metric for measuring a system's ability to achieve a functional goal with high accuracy (i,e task success rate and dialogue cost)~\cite{Rastogi_Zang_Sunkara_Gupta_Khaitan_2020}. Unlike \ac{TDS}, the main focus in chat-bot evaluation is on the user experience during interaction (i.e how engaging, interesting etc.\ the system is) \cite{Li2019ACUTEEVALID}.

The metrics proposed in \citep{gopalakrishnan19_interspeech,Venkatesh2018OnEA} provide a granular analysis on how they influence user satisfaction for chat-bots -- but it is not known how these aspects influence user satisfaction of \acp{TDS}~\citep[see, e.g.,][]{10.1145/3394486.3403202,10.1145/2911451.2911521}. In this study, we focus on understanding the significance of several dialogue aspects on overall impression rating of a \ac{TDS}. We investigate some of the metrics from \citep{gopalakrishnan19_interspeech} originally introduced for chat-bots (viz.\ \emph{interestingness}, \emph{relevance}, and \emph{understanding}) and how they determine a user's overall impression of a \ac{TDS}. 
We also propose a new aspect, \emph{interest arousal}, as a metric in measuring \ac{TDS} effectiveness. 
We find that this newly proposed metric achieves the highest correlation with overall \emph{user satisfaction} with a \ac{TDS}, compared to other metrics that focus on the user experience, with a Spearman's $\rho$ of 0.7903.

To understand the influence of the dialogue aspects, we collect human quality annotations for the \ac{ReDial} dataset \citep{li2018conversational}. The dialogues are annotated at both the turn and dialogue levels on several aspects stated above and extensive analysis is conducted on the annotated dataset. With these we sought to answer the following questions:
What dialogue aspects influence overall impression of \acp{TDS}? and;
What role does the utility and user experience dimensions play when rating the overall impression of a task oriented dialogue?

The contributions we make in this paper are: 
\begin{enumerate*}[label=(\roman*)]
  \item We add an extra annotation layer for the \ac{ReDial} dataset. A human quality annotation effort is set up on \ac{AMT},\footnote{\url{https://mturk.com}} for the annotation of 40 sampled dialogues at the  turn and dialogue level on six dialogue aspects: \textit{relevance}, \textit{interestingness}, \textit{understanding}, \textit{task completion}, \textit{efficiency}, and \textit{interest arousal}.
  \item We analyse the annotated dataset to identify dialogue aspects that influence the overall impression.
  \item We propose additional dialogue aspects with significant contributions to the overall impression of a \ac{TDS}. We classify the annotators' open comments left in the justification box into different categories. Apart from the six dialogue aspects investigated in this study, \emph{natural conversation}, \emph{success in the last interaction} and \emph{repetition} are among other aspects stated by the annotators that influenced their overall impression.
 
\end{enumerate*}
\section{Related Work}
\textbf{User satisfaction.}
 \citet{10.1561/1500000012} defines \textit{user satisfaction} as the fulfilment of a user's specified desire or goal.
User satisfaction has gained popularity as an evaluation metric of \ac{IR} systems based on implicit signals~\citep{10.1145/2556195.2556220,10.1145/2684822.2685319,10.1145/2911451.2911521,10.1145/2854946.2854961,10.1145/3269206.3271802}. Due to the reliance on the user's intelligence and emotions to measure user satisfaction, user satisfaction in information systems depends on the user's interaction experience and the fulfilment of their specified desires, and goals \cite{10.1561/1500000012}.
Factors such as system effectiveness, user effort, user characteristics and expectations influence a user's satisfaction rating for \ac{IR} systems \citep{al2010review}. 
In \ac{TDS}, user satisfaction is measured by rating a dialogue at both the turn and dialogue level on overall impression \cite{10.1145/3404835.3463241,10.1145/3340631.3394856}. 
We rate both the turn and dialogue level with fine-grained aspects of user satisfaction in our work.
 
\header{Dialogue qualities}
Dialogue systems are often evaluated on their overall impression~\citep{Deriu2020SurveyOE}. Recently, research into fine-grained user satisfaction has increased.
\citet{10.3115/976909.979652} propose a framework for evaluating dialogues in a multi-faceted way. It measures several dialogue qualities and combines them to estimate user satisfaction. \citet{mehri-eskenazi-2020-usr} develop an automatic evaluation metric for evaluating dialogue systems at a fine-grained level, including interestingness, engagingness, diversity, understanding, specificity, and inquisitiveness.
Several other publications have investigated human evaluation of dialogue systems on different dialogue qualities~\citep[see, e.g.,][]{dziri-etal-2019-evaluating,see-etal-2019-makes}. 
Our work rates user satisfaction at both turn and dialogue level on six fine-grained user satisfaction aspects, unlike previous research rating both levels on overall impression. We also propose a new aspect, \emph{interest arousal}, which strongly correlates with the overall impression.

\section{Methodology}
To establish dialogue aspects that lead to overall user satisfaction with a \ac{TDS}, we create an additional annotation layer for the \ac{ReDial} dataset. 
We set up an annotation experiment on \acf{AMT} using so-called master workers.
The \ac{AMT}  master workers annotate a total of $40$ conversations on six dialogue aspects. Following \citet{mehri-eskenazi-2020-usr} work, we hand-selected three system responses from each conversation for turn-level annotation. Each response has two previous turns as context plus the following user utterance. Unlike \citep{10.1145/3404835.3463241}, we ask the annotators to decide on the label by considering the user's next action. We display all three turns on a single page and instruct the annotators to answer questions for each turn. After completing the turn-level annotation, the same annotators are taken to a new page where they provide dialogue-level annotations on the same dialogue. The annotators cannot return to the turn-level annotation page. This restriction is based on two considerations:
\begin{enumerate*}[label=(\roman*)]
\item to avoid bias of annotators on the turn-level labels when making decisions on the dialogue-level annotations; and
\item to prevent annotators from going back to change their turn-level ratings.
\end{enumerate*} With this we aim to capture how well an annotator's turn ratings correlate with their dialogue-level ratings and overall ratings.
 
We crowd-source the annotations to enable scalable and efficient annotation labels while capturing different points of view. 
We refined the instructions in a series of pilot studies and internal reviews to ensure the workers understood the task. Moreover, we clearly define each evaluation aspect, backed by real examples from the dataset. 
In the instructions, we stress the fact that the workers need to base their judgements on evidence present in the dialogue (e.g., ``I really liked your suggestion.'') to show relevance, not their personal taste or guess.
We annotate each dialogue with $5$ workers. The annotators answered two questions for each turn and five at the dialogue level.
In total, the annotated dataset includes 1,200 turn-level and 1000 dialogue-level data points. 
At the end of each dialogue, we ask each annotator to leave an open comment to justify their overall impression rating of the \ac{TDS}. 
We obtain 200 open comments from $32$ workers, which we use for analysis to propose additional aspects to be studied with respect user satisfaction as shown in Table~\ref{tab:addaspecs}.

\header{Recommendation dialogue dataset}
The \ac{ReDial} dataset \cite{li2018conversational} is a large dialogue-based human-human movie recommendation corpus. It consists of $11,348$ dialogues. One person is the movie seeker, and the other is the recommender. The movie seeker should explain the kind of movie they like and ask for suggestions. The recommender tries to understand the seeker's movie tastes and recommends movies. This dataset is categorised as both chit-chat and task-oriented since the recommender needs to discover the seeker's movie preference before recommending.

\header{Turn-level annotation}
At the turn level, given the previous context, and the next user utterance, we instruct the workers to assess the system's response according to two fine-grained aspects on a scale of $1$ to $3$:
\begin{description}[leftmargin=\parindent,nosep]
  \item[\rm\textit{Relevance (1–3):}] The system's response is appropriate to the previous turn and fulfils the user's interest ~\citep{see-etal-2019-makes,gopalakrishnan19_interspeech,mehri-eskenazi-2020-usr,shin2020generating}.
  \item[\rm\textit{Interestingness (1–3):}] The system makes chit-chat while presenting facts ~\cite{see-etal-2019-makes,gopalakrishnan19_interspeech,mehri-eskenazi-2020-usr,sun-etal-2021-adding}.
\end{description}
The options for each aspect were: \textit{No, Somewhat, Yes}. For \textit{Relevance}, we also provided a \textit{Not applicable} option in case an annotator believes there is not enough evidence to determine whether the response is relevant (e.g., if no movie is recommended).
 
\header{Dialogue-level annotation}
The workers rate the system's quality considering the entire dialogue for the dialogue-level annotation. 
This level is evaluated based on four aspects. For \textit{understanding, task completion, and interest arousal} we provide three options: \textit{No, Somewhat, Yes}. For \textit{efficiency} raters are given a binary choice~\citep{10.1145/2854946.2854961, EDLUND2008630}. 
We also ask workers to rate the system on overall impression. 
The \textit{overall impression} is rated on a Likert scale of $1$ to $5$ (with $1$ being very dissatisfied and $5$ very satisfied). Below is a summary of the definitions we provide our workers for the dialogue aspects:
\begin{description}[leftmargin=\parindent,nosep]
  \item[\rm\textit{Understanding (1–3):}] The system understands the user's request and fulfils~\citep{Venkatesh2018OnEA, mehri-eskenazi-2020-usr, EDLUND2008630}.
  \item[\rm\textit{Task completion (1–3):}] The system makes suggestions that the user finally accepts~\citep{10.1145/2854946.2854961}.
  \item[\rm\textit{Efficiency (0–1):}] The system can make suggestions that meet the user's interest within the first three interactions~\citep{10.1145/2854946.2854961,EDLUND2008630}.
  \item[\rm\textit{Interest arousal (1–3):}] The system attempts to intrigue the user's interest into accepting a suggestion they are not familiar with.
  \item[\rm\textit{Overall impression (1–5):}] The worker's overall impression of the system's performance, given the dialogue context~\citep{see-etal-2019-makes,gopalakrishnan19_interspeech,mehri-eskenazi-2020-usr,10.1145/2854946.2854961,sun-etal-2021-adding}.
\end{description}

\noindent
Finally, we ask the workers to justify their rating on \emph{overall impression}. We use the justifications to contextualise the given ratings and analyse and discover additional aspects that affect the quality of a dialogue.

\header{Participants}
A total of 32 \ac{AMT} workers took part in the human annotation effort, 18 female and 14 male. Their ages range from 18 to 49.
\begin{table}[!t]
  \caption{Correlation of overall impression with turn-level and dialogue-level annotations. All correlations in this table are statistically significant ($p < 0.01$).}
  \label{tab:turn-level}
  \begin{tabular}{ ll ccc}
 \toprule
 Level & Aspect & Spearman's $\rho$    & Pearson's $r$  \\
 \midrule
 \multirow{2}{*}{Turn} & Relevance & \textbf{0.5199} & \textbf{0.5622} \\
 & Interestingness & 0.3374 & 0.3603 \\
 \midrule
 \multirow{4}{*}{Dialogue} &Understanding& 0.7589 & 0.7928\\
 & Task completion& 0.7895 & 0.8280\\
 & Interest arousal & \textbf{0.7903} & \textbf{0.8341} \\
 & Efficiency & 0.5946& 0.5697\\
 \bottomrule
 \end{tabular}
\end{table}

\section{Results and Analysis}
This section presents the results from our annotation effort and an analysis of the annotators' comments on their overall impression labels. We intend to answer the following questions:
\begin{enumerate*}[label=(RQ\arabic*)]
  \item To what extent do turn-level aspects correlate with the overall user impression in \ac{TDS}? And
  \item What dialogue-level aspects have a more significant influence on the overall user impression?
\end{enumerate*}

As explained above, apart from rating dialogues on six dialogue aspects, annotators also rated the system on overall impression. We use these ratings to classify the dialogues into several categories.
First, a dialogue is \emph{satisfactory} if it has a majority rating of 3 or more; it is \emph{unsatisfactory} if it has a majority rating of less than 3.
Second, we label a dialogue as being \emph{subjective} if: 
\begin{enumerate*}[label=(\roman*)]
   \item two or more annotators selected labels that indicated both satisfactory and unsatisfactory, and
   \item only two annotators agreed on a label whereas the other three selected different labels from each other. That is, we have four different labels selected.
\end{enumerate*}
There are 26 \emph{satisfactory} and 6 \emph{unsatisfactory} dialogues; 8 dialogues are categorised as \emph{subjective}. Inter annotator agreement for the overall impression ratings was fair, with a Fleiss Kappa score of 0.412.

\subsection{Turn-level aspects influencing overall impression}
We compute Spearman's $\rho$ and Pearson's $r$ correlation coefficients with the overall impression for each turn and average across the three turns; see Table~\ref{tab:turn-level} (top). 
The \textit{relevance} aspect exhibits the highest correlation at the turn-level. Out of the dialogues classified as satisfactory, $46\%$ of the turns were rated relevant ($ = 3$) compared to $31\%$ rated interesting ($ = 3$). Hence, more system responses are found to be relevant but not interesting as a \ac{TDS} is traditionally expected to optimise towards task success and not engagement.

When a turn is relevant, the dialogue's overall impression is more likely to be satisfactory ($96\%$ of the turns). The same does not hold for a nonrelevant turn ($43\%$ of the turns led to a satisfactory dialogue), suggesting that in this case, the user's overall impression depends not only on \emph{relevance} but on other dialogue aspects too. A system's success in making a successful suggestion\footnote{A successful suggestion is a movie suggestion that the user accepts.} in the final turn has more weight on the overall impression than the preceding turns. This conforms to the findings of \citep{10.1145/2854946.2854961,liu2020investigating}, which shows that the latest interactions with a system more influence the overall satisfaction of users.
Although relevance is essential in determining the overall impression, it is not the only influencing aspect.

\begin{table}[!t]
  \caption{Determinant coefficients computed with regression showing the effect size of all aspects to overall impression. }
  \label{tab:signifiance}
\setlength{\tabcolsep}{1.25mm}
  \begin{tabular}{ ll @{} cc c}
 \toprule
 & \bf Aspect & \bf Utility & \bf User experience & $R^2$  \\
 \midrule
 \multirow{3}{*}{\rotatebox[origin=c]{90}{Turn~(T)}} & Relevance (R) & $+$ & & 0.568 \\
 & Interestingness (I) & & $+$ & 0.258 \\
& R $+$ I & $+$ & $+$ & \textbf{0.583} \\

 \midrule
 \multirow{5}{*}{\rotatebox[origin=c]{90}{Dialogue~(D)}} & Understanding (U) & & $+$ & 0.629 \\
 & Task completion (TC)& $+$ & & 0.686 \\
 & Interest arousal~(IA)& & $+$ & 0.696 \\
 & Efficiency~(E)& & $+$ & 0.325 \\
 & IA $+$ TC $+$ U $+$ E & $+$ & $+$ &  \textbf{0.825}  \\
 \midrule
 \multirow{3}{*}{\rotatebox[origin=c]{90}{D $+$ T}} & R $+$ TC & $+$ & & 0.761 \\
 & IA $+$ U $+$ I $+$ E & & $+$ & 0.803\\
 & IA $+$ TC $+$ U $+$ I $+$ E $+$ R & $+$ & $+$ & \textbf{0.844}  \\
 \bottomrule
 \end{tabular}
\end{table}

\begin{table*}[h]
%
  \caption{Additional aspects captured from the open comments. The $\%$ show how often the aspect was stated.}
  \label{tab:addaspecs}
  \begin{tabular}{ p{0.18\textwidth} p{0.36\textwidth} p{0.36\textwidth} }
 \toprule
\bf Aspect & \bf Definition & \bf Annotator comment \\
 \midrule
 Opinion~($2.4\%$) & System expresses general opinions on a generic topic or expressing strong personal opinion & ``I don't think that the system should be providing its own opinions on the movies'' \\
 \midrule
 Naturalness~($5.42\%$) & The flow of the conversation is good and fluent & ``The conversation flow naturally from one exchange to the next'' \\
 \midrule
 \raggedright Success on the last interaction~($10.8\%$) & System gets better as time goes by & 
 ``The system finally recommends a good movie at the very end''\\
 \midrule
 Repetition~($1.8\%$)& The system repeating itself or suggestions & ``The system has good suggestions, but it repeats itself over and over which is strange.''\\
 \midrule
 User~($4.21\%$)& User's actions influencing the overall impression & ``The system was being helpful but the user was difficult in answering preference questions'' \\
 \bottomrule
 \end{tabular}
\end{table*}

\subsection{Dialogue-level aspects with significant influence on overall impression}
\begin{figure}[!b]
    \centering
    \subfloat[]{\includegraphics[width=0.5\columnwidth]{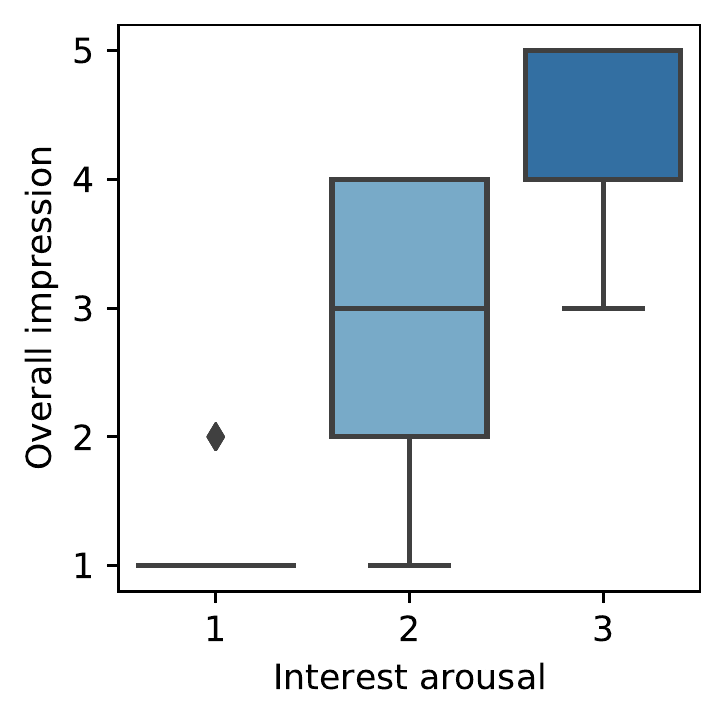}\label{fig:diagaspects:interest}}
    \subfloat[]{\includegraphics[width=0.5\columnwidth]{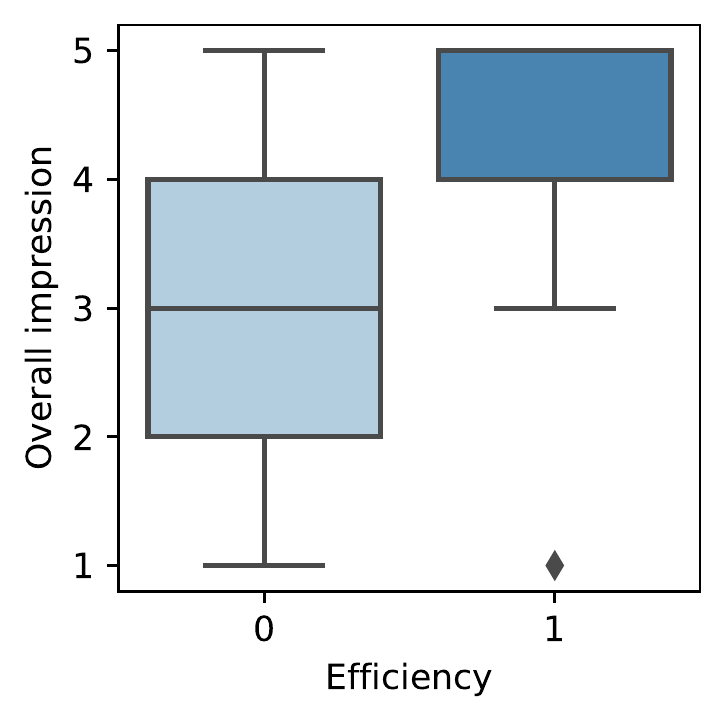}\label{fig:diagaspects:efficiency}}
    \caption{Box plots showing distribution  of the (a) \emph{interest arousal} and (b) \emph{efficiency} aspects ratings against overall impression ratings.}
    \label{fig:diagaspects}
\end{figure}

In Figure~\ref{fig:diagaspects} we plot the distribution of the ratings for the dialogue-level aspects against overall impression. We see a clear dependency of the \textit{overall impression} on the \textit{interest arousal} aspect; out of the dialogues classified as satisfactory, $73\%$ were rated high in terms of interest arousal (see Figure~\ref{fig:diagaspects:interest}). We also notice that all dialogues rated low ($=1$) are unsatisfactory overall. Thus the ability of a \ac{TDS} to intrigue the user's interest in watching a novel suggestion can be the determinant of the overall impression.

Table~\ref{tab:turn-level} (bottom) reports Spearman's $\rho$ and Pearson's $r$ correlation coefficients of the dialogue-level aspects with overall impression rating. \emph{Efficiency} is the least correlating aspect for both scores. In our study, this aspect captures the system's ability to make relevant recommendations meeting the user's need within the first three exchanges. Unlike chatbots, which are meant to engage with a user for a long period, \ac{TDS} dialogues should be concise~\citep{gao-2021-advances}.

We see in Figure~\ref{fig:diagaspects:efficiency} that more dialogues are rated inefficient than efficient ($53.5\%$ vs.~$46.5\%$ ), suggesting that efficient suggestions of movies contribute to a dialogue being satisfactory. Our analysis, however, indicates that the opposite cannot be said for inefficient dialogues: some of them were rated satisfactory ($64.44\%$). We note from the annotators' open comments that though a system took extra turns to make a relevant suggestion, as long as the user got a suggestion, they rate the system as satisfactory. 
This indicates that a system that fails to satisfy the user's need in the first three interactions is less likely to do so in further interactions.

To understand the significance of the investigated dialogue aspects to the overall impression, we train various regression models considering different aspect combinations (both single and multiple aspects); see Table~\ref{tab:signifiance} for the results. 
At the turn-level, an approach that combines both aspects outperforms the best turn-level single aspect (\emph{relevance}). As for the dialogue-level aspects, \emph{interest arousal} exhibits the highest significance among all other aspects, taken individually. The combination of dialogue-level aspects clearly shows a stronger relationship to the overall rating model than individual aspects.
Unsurprisingly, combining all aspects performs better than that of individual aspects or different levels. 

Tables~\ref{tab:turn-level} and~\ref{tab:signifiance} show that dialogue-level aspects have a bigger influence on the overall impression than turn-level aspects. This suggests that turn-level aspects cannot be used solely to estimate the user's overall satisfaction effectively. 
This is attributed to cases where a system's response at a turn is sub-optimal, thus not representing the entire dialogue impression. 
The turn and dialogue aspects concern two evaluation dimensions: utility and user experience. 
\emph{Relevance} and \emph{task completion} measure the utility of a \ac{TDS}, i.e., its ability to accomplish a task by making relevant suggestions. The user experience dimensions (\emph{understanding}, \emph{interest arousal}, \emph{efficiency} and \emph{interestingness}) focus on the user's interaction experience. 
Combining dialogue aspects from both dimensions has a strong relationship to the overall impression, unlike the individual aspects. 
In Table~\ref{tab:signifiance} the columns Utility and User experience show the two dimensions: combining both dimensions (the last row in each section in Table~\ref{tab:signifiance}) leads to the best performance.
The combination of turn and dialogue level aspects (D$+$T, third group) achieves the highest $R^2$. 
In summary, leveraging aspects from both dimensions (utility and user experience) is essential when designing a \ac{TDS} that is meant to achieve a high overall impression.

\vspace*{-2mm}
\subsection{Analysis of the justifications}
\label{justificatio}
We report on a manual inspection of the workers' open comments. 
We went through the comments and assigned them to evaluation aspects based on the worker's perspective. E.g., a comment that mentions ``the system kept recommending the same movie'' signals the existence of a novel aspect that concerns repeated recommendations in a dialogue. 
Table~\ref{tab:addaspecs} lists the (dominant) novel categories discovered from the comments, together with a gloss and example. 
Several interesting aspects are observed by the annotators. 
For example, most annotators disliked the fact that the system expressed its opinion on a genre or movie. 
In cases where the system is repetitive (in terms of language use or recommended items), the annotators' assessments were negatively impacted. 
Some annotators noted the positive impact of a dialogue being natural and human-like or that the system made a good recommendation after several failed suggestions (i.e., success on the last interaction). 
There were some examples where all annotators agreed that the suggestions were good, but the user did not react rationally.

\section{Discussion and Conclusions}
In this paper, we focus on providing a fine-grained understanding of what the overall user impression means in \acp{TDS}. We asked annotators to follow a dialogue and assess both at the turn and dialogue level on multiple aspects. While related work highlights the significance of these aspects~\cite{sun-etal-2021-adding,see-etal-2019-makes,zhang-etal-2018-personalizing}, not much work has been done on the impact of these aspects on \acp{TDS}.

Providing relevant recommendations throughout a dialogue is crucial for user satisfaction, but it does not tell the whole story. Both from the annotations and open-ended comments, we find that engaging with users in the form of chit-chat can have two effects. If a user is already happy with a provided recommendation, more engagement can lead to further \textit{interest arousal}, and hence more satisfaction; but if the system fails to meet the user's expectations, it can have a negative effect. This is in line with \citep{sun-etal-2021-adding}, who stress the importance of finding the right amount of chit-chat in a dialogue.

Our analysis of open-ended comments and justifications revealed new aspects that can affect users' satisfaction. In line with our quantitative analysis and related work~\cite{10.1145/2854946.2854961,liu2020investigating}, many annotators mentioned the importance of user experience in the final turns or at least one successful interaction in the dialogue. Other aspects such as repeated utterances and recommendations negatively impacted the user experience. This indicates the need for jointly optimising turn- and dialogue-level metrics and for a fine-grained model of user satisfaction that incorporates multiple aspects.

One limitation of our work is that the annotators assess user satisfaction based on the user's utterances and reactions to the system's responses. While we observed a high level of agreement for most dialogues, we noticed a disagreement between annotators on some dialogues. We plan to collect a set of fine-grained annotation labels directly from users. Also, we plan to extend this study to learn to predict the overall impression of the users in a \ac{TDS}.

\begin{acks}
We thank our reviewers for valuable feedback.
This research was supported by Huawei Finland and by the Hybrid Intelligence Center, a 10-year program funded by the Dutch Ministry of Education, Culture and Science through the Netherlands Organisation for Scientific Research, \url{https://hybrid-intelligence-centre.nl}. 
All content represents the opinion of the authors, which is not necessarily shared or endorsed by their respective employers and/or sponsors.
\end{acks}

\bibliographystyle{ACM-Reference-Format}
\balance
\bibliography{references}


\begin{thebibliography}{27}


\ifx \showCODEN    \undefined \def \showCODEN     #1{\unskip}     \fi
\ifx \showDOI      \undefined \def \showDOI       #1{#1}\fi
\ifx \showISBNx    \undefined \def \showISBNx     #1{\unskip}     \fi
\ifx \showISBNxiii \undefined \def \showISBNxiii  #1{\unskip}     \fi
\ifx \showISSN     \undefined \def \showISSN      #1{\unskip}     \fi
\ifx \showLCCN     \undefined \def \showLCCN      #1{\unskip}     \fi
\ifx \shownote     \undefined \def \shownote      #1{#1}          \fi
\ifx \showarticletitle \undefined \def \showarticletitle #1{#1}   \fi
\ifx \showURL      \undefined \def \showURL       {\relax}        \fi
\providecommand\bibfield[2]{#2}
\providecommand\bibinfo[2]{#2}
\providecommand\natexlab[1]{#1}
\providecommand\showeprint[2][]{arXiv:#2}

\bibitem[\protect\citeauthoryear{Al-Maskari and Sanderson}{Al-Maskari and
  Sanderson}{2010}]%
        {al2010review}
\bibfield{author}{\bibinfo{person}{Azzah Al-Maskari} {and}
  \bibinfo{person}{Mark Sanderson}.} \bibinfo{year}{2010}\natexlab{}.
\newblock \showarticletitle{A review of factors influencing user satisfaction
  in information retrieval}.
\newblock \bibinfo{journal}{\emph{Journal of the American Society for
  Information Science and Technology}} \bibinfo{volume}{61},
  \bibinfo{number}{5} (\bibinfo{year}{2010}), \bibinfo{pages}{859--868}.
\newblock


\bibitem[\protect\citeauthoryear{Cai and Chen}{Cai and Chen}{2020}]%
        {10.1145/3340631.3394856}
\bibfield{author}{\bibinfo{person}{Wanling Cai} {and} \bibinfo{person}{Li
  Chen}.} \bibinfo{year}{2020}\natexlab{}.
\newblock \bibinfo{booktitle}{\emph{Predicting User Intents and Satisfaction
  with Dialogue-Based Conversational Recommendations}}.
\newblock \bibinfo{publisher}{Association for Computing Machinery},
  \bibinfo{address}{New York, NY, USA}, \bibinfo{pages}{33–42}.
\newblock
\showISBNx{9781450368612}
\urldef\tempurl%
\url{https://doi.org/10.1145/3340631.3394856}
\showURL{%
\tempurl}


\bibitem[\protect\citeauthoryear{Deriu, Rodrigo, Otegi, Echegoyen, Rosset,
  Agirre, and Cieliebak}{Deriu et~al\mbox{.}}{2020}]%
        {Deriu2020SurveyOE}
\bibfield{author}{\bibinfo{person}{Jan Deriu}, \bibinfo{person}{{\'A}lvaro
  Rodrigo}, \bibinfo{person}{Arantxa Otegi}, \bibinfo{person}{Guillermo
  Echegoyen}, \bibinfo{person}{Sophie Rosset}, \bibinfo{person}{Eneko Agirre},
  {and} \bibinfo{person}{Mark Cieliebak}.} \bibinfo{year}{2020}\natexlab{}.
\newblock \showarticletitle{Survey on evaluation methods for dialogue systems}.
\newblock \bibinfo{journal}{\emph{Artificial Intelligence Review}}
  \bibinfo{volume}{54} (\bibinfo{year}{2020}), \bibinfo{pages}{755 -- 810}.
\newblock


\bibitem[\protect\citeauthoryear{Dziri, Kamalloo, Mathewson, and Zaiane}{Dziri
  et~al\mbox{.}}{2019}]%
        {dziri-etal-2019-evaluating}
\bibfield{author}{\bibinfo{person}{Nouha Dziri}, \bibinfo{person}{Ehsan
  Kamalloo}, \bibinfo{person}{Kory Mathewson}, {and} \bibinfo{person}{Osmar
  Zaiane}.} \bibinfo{year}{2019}\natexlab{}.
\newblock \showarticletitle{Evaluating Coherence in Dialogue Systems using
  Entailment}. In \bibinfo{booktitle}{\emph{Proceedings of the 2019 Conference
  of the North {A}merican Chapter of the Association for Computational
  Linguistics: Human Language Technologies, Volume 1 (Long and Short Papers)}}.
  \bibinfo{publisher}{Association for Computational Linguistics},
  \bibinfo{address}{Minneapolis, Minnesota}, \bibinfo{pages}{3806--3812}.
\newblock
\urldef\tempurl%
\url{https://doi.org/10.18653/v1/N19-1381}
\showDOI{\tempurl}


\bibitem[\protect\citeauthoryear{Edlund, Gustafson, Heldner, and
  Hjalmarsson}{Edlund et~al\mbox{.}}{2008}]%
        {EDLUND2008630}
\bibfield{author}{\bibinfo{person}{Jens Edlund}, \bibinfo{person}{Joakim
  Gustafson}, \bibinfo{person}{Mattias Heldner}, {and} \bibinfo{person}{Anna
  Hjalmarsson}.} \bibinfo{year}{2008}\natexlab{}.
\newblock \showarticletitle{Towards human-like spoken dialogue systems}.
\newblock \bibinfo{journal}{\emph{Speech Communication}} \bibinfo{volume}{50},
  \bibinfo{number}{8} (\bibinfo{year}{2008}), \bibinfo{pages}{630--645}.
\newblock
\showISSN{0167-6393}
\urldef\tempurl%
\url{https://doi.org/10.1016/j.specom.2008.04.002}
\showDOI{\tempurl}
\newblock
\shownote{Evaluating new methods and models for advanced speech-based
  interactive systems}.


\bibitem[\protect\citeauthoryear{Gao, Lei, He, de~Rijke, and Chua}{Gao
  et~al\mbox{.}}{2021}]%
        {gao-2021-advances}
\bibfield{author}{\bibinfo{person}{Chongming Gao}, \bibinfo{person}{Wenqiang
  Lei}, \bibinfo{person}{Xiangnan He}, \bibinfo{person}{Maarten de Rijke},
  {and} \bibinfo{person}{Tat-Seng Chua}.} \bibinfo{year}{2021}\natexlab{}.
\newblock \showarticletitle{Advances and Challenges in Conversational
  Recommender Systems: A Survey}.
\newblock \bibinfo{journal}{\emph{AI Open}}  \bibinfo{volume}{2}
  (\bibinfo{date}{July} \bibinfo{year}{2021}), \bibinfo{pages}{100--126}.
\newblock


\bibitem[\protect\citeauthoryear{Gopalakrishnan, Hedayatnia, Chen, Gottardi,
  Kwatra, Venkatesh, Gabriel, and Hakkani-Tür}{Gopalakrishnan
  et~al\mbox{.}}{2019}]%
        {gopalakrishnan19_interspeech}
\bibfield{author}{\bibinfo{person}{Karthik Gopalakrishnan},
  \bibinfo{person}{Behnam Hedayatnia}, \bibinfo{person}{Qinlang Chen},
  \bibinfo{person}{Anna Gottardi}, \bibinfo{person}{Sanjeev Kwatra},
  \bibinfo{person}{Anu Venkatesh}, \bibinfo{person}{Raefer Gabriel}, {and}
  \bibinfo{person}{Dilek Hakkani-Tür}.} \bibinfo{year}{2019}\natexlab{}.
\newblock \showarticletitle{{Topical-Chat: Towards Knowledge-Grounded
  Open-Domain Conversations}}. In \bibinfo{booktitle}{\emph{Proc. Interspeech
  2019}}. \bibinfo{pages}{1891--1895}.
\newblock
\urldef\tempurl%
\url{https://doi.org/10.21437/Interspeech.2019-3079}
\showDOI{\tempurl}


\bibitem[\protect\citeauthoryear{Hashemi, Williams, El~Kholy, Zitouni, and
  Crook}{Hashemi et~al\mbox{.}}{2018}]%
        {10.1145/3269206.3271802}
\bibfield{author}{\bibinfo{person}{Seyyed~Hadi Hashemi}, \bibinfo{person}{Kyle
  Williams}, \bibinfo{person}{Ahmed El~Kholy}, \bibinfo{person}{Imed Zitouni},
  {and} \bibinfo{person}{Paul~A. Crook}.} \bibinfo{year}{2018}\natexlab{}.
\newblock \showarticletitle{Measuring User Satisfaction on Smart Speaker
  Intelligent Assistants Using Intent Sensitive Query Embeddings}. In
  \bibinfo{booktitle}{\emph{Proceedings of the 27th ACM International
  Conference on Information and Knowledge Management}} (Torino, Italy)
  \emph{(\bibinfo{series}{CIKM '18})}. \bibinfo{publisher}{Association for
  Computing Machinery}, \bibinfo{address}{New York, NY, USA},
  \bibinfo{pages}{1183–1192}.
\newblock
\showISBNx{9781450360142}
\urldef\tempurl%
\url{https://doi.org/10.1145/3269206.3271802}
\showDOI{\tempurl}


\bibitem[\protect\citeauthoryear{Jiang, Hassan~Awadallah, Shi, and White}{Jiang
  et~al\mbox{.}}{2015}]%
        {10.1145/2684822.2685319}
\bibfield{author}{\bibinfo{person}{Jiepu Jiang}, \bibinfo{person}{Ahmed
  Hassan~Awadallah}, \bibinfo{person}{Xiaolin Shi}, {and}
  \bibinfo{person}{Ryen~W. White}.} \bibinfo{year}{2015}\natexlab{}.
\newblock \showarticletitle{Understanding and Predicting Graded Search
  Satisfaction}. In \bibinfo{booktitle}{\emph{Proceedings of the Eighth ACM
  International Conference on Web Search and Data Mining}} (Shanghai, China)
  \emph{(\bibinfo{series}{WSDM '15})}. \bibinfo{publisher}{Association for
  Computing Machinery}, \bibinfo{address}{New York, NY, USA},
  \bibinfo{pages}{57–66}.
\newblock
\showISBNx{9781450333177}
\urldef\tempurl%
\url{https://doi.org/10.1145/2684822.2685319}
\showDOI{\tempurl}


\bibitem[\protect\citeauthoryear{Kelly}{Kelly}{2009}]%
        {10.1561/1500000012}
\bibfield{author}{\bibinfo{person}{Diane Kelly}.}
  \bibinfo{year}{2009}\natexlab{}.
\newblock \showarticletitle{Methods for Evaluating Interactive Information
  Retrieval Systems with Users}.
\newblock \bibinfo{journal}{\emph{Found. Trends Inf. Retr.}}
  \bibinfo{volume}{3}, \bibinfo{number}{1—2} (\bibinfo{date}{jan}
  \bibinfo{year}{2009}), \bibinfo{pages}{1–224}.
\newblock
\showISSN{1554-0669}
\urldef\tempurl%
\url{https://doi.org/10.1561/1500000012}
\showDOI{\tempurl}


\bibitem[\protect\citeauthoryear{Kim, Hassan, White, and Zitouni}{Kim
  et~al\mbox{.}}{2014}]%
        {10.1145/2556195.2556220}
\bibfield{author}{\bibinfo{person}{Youngho Kim}, \bibinfo{person}{Ahmed
  Hassan}, \bibinfo{person}{Ryen~W. White}, {and} \bibinfo{person}{Imed
  Zitouni}.} \bibinfo{year}{2014}\natexlab{}.
\newblock \showarticletitle{Modeling Dwell Time to Predict Click-Level
  Satisfaction}. In \bibinfo{booktitle}{\emph{Proceedings of the 7th ACM
  International Conference on Web Search and Data Mining}} (New York, New York,
  USA) \emph{(\bibinfo{series}{WSDM '14})}. \bibinfo{publisher}{Association for
  Computing Machinery}, \bibinfo{address}{New York, NY, USA},
  \bibinfo{pages}{193–202}.
\newblock
\showISBNx{9781450323512}
\urldef\tempurl%
\url{https://doi.org/10.1145/2556195.2556220}
\showDOI{\tempurl}


\bibitem[\protect\citeauthoryear{Kiseleva, Williams, Hassan~Awadallah, Crook,
  Zitouni, and Anastasakos}{Kiseleva et~al\mbox{.}}{2016a}]%
        {10.1145/2911451.2911521}
\bibfield{author}{\bibinfo{person}{Julia Kiseleva}, \bibinfo{person}{Kyle
  Williams}, \bibinfo{person}{Ahmed Hassan~Awadallah},
  \bibinfo{person}{Aidan~C. Crook}, \bibinfo{person}{Imed Zitouni}, {and}
  \bibinfo{person}{Tasos Anastasakos}.} \bibinfo{year}{2016}\natexlab{a}.
\newblock \showarticletitle{Predicting User Satisfaction with Intelligent
  Assistants}. In \bibinfo{booktitle}{\emph{Proceedings of the 39th
  International ACM SIGIR Conference on Research and Development in Information
  Retrieval}} (Pisa, Italy) \emph{(\bibinfo{series}{SIGIR '16})}.
  \bibinfo{publisher}{Association for Computing Machinery},
  \bibinfo{address}{New York, NY, USA}, \bibinfo{pages}{45–54}.
\newblock
\showISBNx{9781450340694}
\urldef\tempurl%
\url{https://doi.org/10.1145/2911451.2911521}
\showDOI{\tempurl}


\bibitem[\protect\citeauthoryear{Kiseleva, Williams, Jiang, Hassan~Awadallah,
  Crook, Zitouni, and Anastasakos}{Kiseleva et~al\mbox{.}}{2016b}]%
        {10.1145/2854946.2854961}
\bibfield{author}{\bibinfo{person}{Julia Kiseleva}, \bibinfo{person}{Kyle
  Williams}, \bibinfo{person}{Jiepu Jiang}, \bibinfo{person}{Ahmed
  Hassan~Awadallah}, \bibinfo{person}{Aidan~C. Crook}, \bibinfo{person}{Imed
  Zitouni}, {and} \bibinfo{person}{Tasos Anastasakos}.}
  \bibinfo{year}{2016}\natexlab{b}.
\newblock \showarticletitle{Understanding User Satisfaction with Intelligent
  Assistants}. In \bibinfo{booktitle}{\emph{Proceedings of the 2016 ACM on
  Conference on Human Information Interaction and Retrieval}} (Carrboro, North
  Carolina, USA) \emph{(\bibinfo{series}{CHIIR '16})}.
  \bibinfo{publisher}{Association for Computing Machinery},
  \bibinfo{address}{New York, NY, USA}, \bibinfo{pages}{121–130}.
\newblock
\showISBNx{9781450337519}
\urldef\tempurl%
\url{https://doi.org/10.1145/2854946.2854961}
\showDOI{\tempurl}


\bibitem[\protect\citeauthoryear{Li, Weston, and Roller}{Li
  et~al\mbox{.}}{2019}]%
        {Li2019ACUTEEVALID}
\bibfield{author}{\bibinfo{person}{Margaret Li}, \bibinfo{person}{Jason
  Weston}, {and} \bibinfo{person}{Stephen Roller}.}
  \bibinfo{year}{2019}\natexlab{}.
\newblock \showarticletitle{ACUTE-EVAL: Improved Dialogue Evaluation with
  Optimized Questions and Multi-turn Comparisons}.
\newblock \bibinfo{journal}{\emph{ArXiv}}  \bibinfo{volume}{abs/1909.03087}
  (\bibinfo{year}{2019}).
\newblock


\bibitem[\protect\citeauthoryear{Li, Kahou, Schulz, Michalski, Charlin, and
  Pal}{Li et~al\mbox{.}}{2018}]%
        {li2018conversational}
\bibfield{author}{\bibinfo{person}{Raymond Li},
  \bibinfo{person}{Samira~Ebrahimi Kahou}, \bibinfo{person}{Hannes Schulz},
  \bibinfo{person}{Vincent Michalski}, \bibinfo{person}{Laurent Charlin}, {and}
  \bibinfo{person}{Chris Pal}.} \bibinfo{year}{2018}\natexlab{}.
\newblock \showarticletitle{Towards Deep Conversational Recommendations}. In
  \bibinfo{booktitle}{\emph{Advances in Neural Information Processing Systems
  31 (NIPS 2018)}}.
\newblock


\bibitem[\protect\citeauthoryear{Liu, Lowe, Serban, Noseworthy, Charlin, and
  Pineau}{Liu et~al\mbox{.}}{2016}]%
        {liu-etal-2016-evaluate}
\bibfield{author}{\bibinfo{person}{Chia-Wei Liu}, \bibinfo{person}{Ryan Lowe},
  \bibinfo{person}{Iulian Serban}, \bibinfo{person}{Mike Noseworthy},
  \bibinfo{person}{Laurent Charlin}, {and} \bibinfo{person}{Joelle Pineau}.}
  \bibinfo{year}{2016}\natexlab{}.
\newblock \showarticletitle{How {NOT} To Evaluate Your Dialogue System: An
  Empirical Study of Unsupervised Evaluation Metrics for Dialogue Response
  Generation}. In \bibinfo{booktitle}{\emph{Proceedings of the 2016 Conference
  on Empirical Methods in Natural Language Processing}}.
  \bibinfo{publisher}{Association for Computational Linguistics},
  \bibinfo{address}{Austin, Texas}, \bibinfo{pages}{2122--2132}.
\newblock
\urldef\tempurl%
\url{https://doi.org/10.18653/v1/D16-1230}
\showDOI{\tempurl}


\bibitem[\protect\citeauthoryear{Liu and Han}{Liu and Han}{2020}]%
        {liu2020investigating}
\bibfield{author}{\bibinfo{person}{Jiqun Liu} {and} \bibinfo{person}{Fangyuan
  Han}.} \bibinfo{year}{2020}\natexlab{}.
\newblock \showarticletitle{Investigating Reference Dependence Effects on User
  Search Interaction and Satisfaction: {A} Behavioral Economics Perspective}.
  In \bibinfo{booktitle}{\emph{Proceedings of the 43rd International {ACM}
  {SIGIR} conference on research and development in Information Retrieval,
  {SIGIR}}}. \bibinfo{publisher}{{ACM}}, \bibinfo{pages}{1141--1150}.
\newblock


\bibitem[\protect\citeauthoryear{Mehri and Eskenazi}{Mehri and
  Eskenazi}{2020}]%
        {mehri-eskenazi-2020-usr}
\bibfield{author}{\bibinfo{person}{Shikib Mehri} {and} \bibinfo{person}{Maxine
  Eskenazi}.} \bibinfo{year}{2020}\natexlab{}.
\newblock \showarticletitle{{USR}: An Unsupervised and Reference Free
  Evaluation Metric for Dialog Generation}. In
  \bibinfo{booktitle}{\emph{Proceedings of the 58th Annual Meeting of the
  Association for Computational Linguistics}}. \bibinfo{publisher}{Association
  for Computational Linguistics}, \bibinfo{address}{Online},
  \bibinfo{pages}{681--707}.
\newblock
\urldef\tempurl%
\url{https://doi.org/10.18653/v1/2020.acl-main.64}
\showDOI{\tempurl}


\bibitem[\protect\citeauthoryear{Rastogi, Zang, Sunkara, Gupta, and
  Khaitan}{Rastogi et~al\mbox{.}}{2020}]%
        {Rastogi_Zang_Sunkara_Gupta_Khaitan_2020}
\bibfield{author}{\bibinfo{person}{Abhinav Rastogi}, \bibinfo{person}{Xiaoxue
  Zang}, \bibinfo{person}{Srinivas Sunkara}, \bibinfo{person}{Raghav Gupta},
  {and} \bibinfo{person}{Pranav Khaitan}.} \bibinfo{year}{2020}\natexlab{}.
\newblock \showarticletitle{Towards Scalable Multi-Domain Conversational
  Agents: The Schema-Guided Dialogue Dataset}.
\newblock \bibinfo{journal}{\emph{Proceedings of the AAAI Conference on
  Artificial Intelligence}} \bibinfo{volume}{34}, \bibinfo{number}{05}
  (\bibinfo{date}{Apr.} \bibinfo{year}{2020}), \bibinfo{pages}{8689--8696}.
\newblock
\urldef\tempurl%
\url{https://doi.org/10.1609/aaai.v34i05.6394}
\showDOI{\tempurl}


\bibitem[\protect\citeauthoryear{See, Roller, Kiela, and Weston}{See
  et~al\mbox{.}}{2019}]%
        {see-etal-2019-makes}
\bibfield{author}{\bibinfo{person}{Abigail See}, \bibinfo{person}{Stephen
  Roller}, \bibinfo{person}{Douwe Kiela}, {and} \bibinfo{person}{Jason
  Weston}.} \bibinfo{year}{2019}\natexlab{}.
\newblock \showarticletitle{What makes a good conversation? How controllable
  attributes affect human judgments}. In \bibinfo{booktitle}{\emph{Proceedings
  of the 2019 Conference of the North {A}merican Chapter of the Association for
  Computational Linguistics: Human Language Technologies, Volume 1 (Long and
  Short Papers)}}. \bibinfo{publisher}{Association for Computational
  Linguistics}, \bibinfo{address}{Minneapolis, Minnesota},
  \bibinfo{pages}{1702--1723}.
\newblock
\urldef\tempurl%
\url{https://doi.org/10.18653/v1/N19-1170}
\showDOI{\tempurl}


\bibitem[\protect\citeauthoryear{Shin, Xu, Madotto, and Fung}{Shin
  et~al\mbox{.}}{2020}]%
        {shin2020generating}
\bibfield{author}{\bibinfo{person}{Jamin Shin}, \bibinfo{person}{Peng Xu},
  \bibinfo{person}{Andrea Madotto}, {and} \bibinfo{person}{Pascale Fung}.}
  \bibinfo{year}{2020}\natexlab{}.
\newblock \showarticletitle{Generating empathetic responses by looking ahead
  the user’s sentiment}. In \bibinfo{booktitle}{\emph{ICASSP 2020-2020 IEEE
  International Conference on Acoustics, Speech and Signal Processing
  (ICASSP)}}. IEEE, \bibinfo{pages}{7989--7993}.
\newblock


\bibitem[\protect\citeauthoryear{Sun, Moon, Crook, Roller, Silvert, Liu, Wang,
  Liu, Cho, and Cardie}{Sun et~al\mbox{.}}{2021a}]%
        {sun-etal-2021-adding}
\bibfield{author}{\bibinfo{person}{Kai Sun}, \bibinfo{person}{Seungwhan Moon},
  \bibinfo{person}{Paul Crook}, \bibinfo{person}{Stephen Roller},
  \bibinfo{person}{Becka Silvert}, \bibinfo{person}{Bing Liu},
  \bibinfo{person}{Zhiguang Wang}, \bibinfo{person}{Honglei Liu},
  \bibinfo{person}{Eunjoon Cho}, {and} \bibinfo{person}{Claire Cardie}.}
  \bibinfo{year}{2021}\natexlab{a}.
\newblock \showarticletitle{Adding Chit-Chat to Enhance Task-Oriented
  Dialogues}. In \bibinfo{booktitle}{\emph{Proceedings of the 2021 Conference
  of the North American Chapter of the Association for Computational
  Linguistics: Human Language Technologies}}. \bibinfo{publisher}{Association
  for Computational Linguistics}, \bibinfo{address}{Online},
  \bibinfo{pages}{1570--1583}.
\newblock
\urldef\tempurl%
\url{https://doi.org/10.18653/v1/2021.naacl-main.124}
\showDOI{\tempurl}


\bibitem[\protect\citeauthoryear{Sun, Zhang, Balog, Ren, Ren, Chen, and
  de~Rijke}{Sun et~al\mbox{.}}{2021b}]%
        {10.1145/3404835.3463241}
\bibfield{author}{\bibinfo{person}{Weiwei Sun}, \bibinfo{person}{Shuo Zhang},
  \bibinfo{person}{Krisztian Balog}, \bibinfo{person}{Zhaochun Ren},
  \bibinfo{person}{Pengjie Ren}, \bibinfo{person}{Zhumin Chen}, {and}
  \bibinfo{person}{Maarten de Rijke}.} \bibinfo{year}{2021}\natexlab{b}.
\newblock \bibinfo{booktitle}{\emph{Simulating User Satisfaction for the
  Evaluation of Task-Oriented Dialogue Systems}}.
\newblock \bibinfo{publisher}{Association for Computing Machinery},
  \bibinfo{address}{New York, NY, USA}, \bibinfo{pages}{2499–2506}.
\newblock
\showISBNx{9781450380379}
\urldef\tempurl%
\url{https://doi.org/10.1145/3404835.3463241}
\showURL{%
\tempurl}


\bibitem[\protect\citeauthoryear{Venkatesh, Khatri, Ram, Guo, Gabriel, Nagar,
  Prasad, Cheng, Hedayatnia, Metallinou, Goel, Yang, and Raju}{Venkatesh
  et~al\mbox{.}}{2018}]%
        {Venkatesh2018OnEA}
\bibfield{author}{\bibinfo{person}{Anu Venkatesh}, \bibinfo{person}{Chandra
  Khatri}, \bibinfo{person}{Ashwin Ram}, \bibinfo{person}{Fenfei Guo},
  \bibinfo{person}{Raefer Gabriel}, \bibinfo{person}{Ashish Nagar},
  \bibinfo{person}{Rohit Prasad}, \bibinfo{person}{Ming Cheng},
  \bibinfo{person}{Behnam Hedayatnia}, \bibinfo{person}{Angeliki Metallinou},
  \bibinfo{person}{Rahul Goel}, \bibinfo{person}{Shaohua Yang}, {and}
  \bibinfo{person}{Anirudh Raju}.} \bibinfo{year}{2018}\natexlab{}.
\newblock \showarticletitle{On Evaluating and Comparing Open Domain Dialog
  Systems}.
\newblock \bibinfo{journal}{\emph{arXiv: Computation and Language}}
  (\bibinfo{year}{2018}).
\newblock


\bibitem[\protect\citeauthoryear{Walker, Litman, Kamm, and Abella}{Walker
  et~al\mbox{.}}{1997}]%
        {10.3115/976909.979652}
\bibfield{author}{\bibinfo{person}{Marilyn~A. Walker},
  \bibinfo{person}{Diane~J. Litman}, \bibinfo{person}{Candace~A. Kamm}, {and}
  \bibinfo{person}{Alicia Abella}.} \bibinfo{year}{1997}\natexlab{}.
\newblock \showarticletitle{PARADISE: A Framework for Evaluating Spoken
  Dialogue Agents}. In \bibinfo{booktitle}{\emph{Proceedings of the 35th Annual
  Meeting of the Association for Computational Linguistics and Eighth
  Conference of the European Chapter of the Association for Computational
  Linguistics}} (Madrid, Spain) \emph{(\bibinfo{series}{ACL '98/EACL '98})}.
  \bibinfo{publisher}{Association for Computational Linguistics},
  \bibinfo{address}{USA}, \bibinfo{pages}{271–280}.
\newblock
\urldef\tempurl%
\url{https://doi.org/10.3115/976909.979652}
\showDOI{\tempurl}


\bibitem[\protect\citeauthoryear{Zhang and Balog}{Zhang and Balog}{2020}]%
        {10.1145/3394486.3403202}
\bibfield{author}{\bibinfo{person}{Shuo Zhang} {and} \bibinfo{person}{Krisztian
  Balog}.} \bibinfo{year}{2020}\natexlab{}.
\newblock \bibinfo{booktitle}{\emph{Evaluating Conversational Recommender
  Systems via User Simulation}}.
\newblock \bibinfo{publisher}{Association for Computing Machinery},
  \bibinfo{address}{New York, NY, USA}, \bibinfo{pages}{1512–1520}.
\newblock
\showISBNx{9781450379984}
\urldef\tempurl%
\url{https://doi.org/10.1145/3394486.3403202}
\showURL{%
\tempurl}


\bibitem[\protect\citeauthoryear{Zhang, Dinan, Urbanek, Szlam, Kiela, and
  Weston}{Zhang et~al\mbox{.}}{2018}]%
        {zhang-etal-2018-personalizing}
\bibfield{author}{\bibinfo{person}{Saizheng Zhang}, \bibinfo{person}{Emily
  Dinan}, \bibinfo{person}{Jack Urbanek}, \bibinfo{person}{Arthur Szlam},
  \bibinfo{person}{Douwe Kiela}, {and} \bibinfo{person}{Jason Weston}.}
  \bibinfo{year}{2018}\natexlab{}.
\newblock \showarticletitle{Personalizing Dialogue Agents: {I} have a dog, do
  you have pets too?}. In \bibinfo{booktitle}{\emph{Proceedings of the 56th
  Annual Meeting of the Association for Computational Linguistics (Volume 1:
  Long Papers)}}. \bibinfo{publisher}{Association for Computational
  Linguistics}, \bibinfo{address}{Melbourne, Australia},
  \bibinfo{pages}{2204--2213}.
\newblock
\urldef\tempurl%
\url{https://doi.org/10.18653/v1/P18-1205}
\showDOI{\tempurl}


\end{thebibliography}

\end{document}